\documentclass[11pt]{GACM2017_0094}
\usepackage{mathtools}
\begin{document}	
\maketitle
\thispagestyle{empty}
		\section{Introduction}
		The LHC (Large Hadron Collider) at CERN, Geneva uses superconducting magnets for focussing and bending particle beams. To ensure their superconductivity, they are cooled to very low temperatures (1.9 K). However, in the case of a malfunction the superconducting material becomes resistive and quenches. The quenched area quickly overheats, the heat propagates within the magnet and from one magnet to another causing more magnets to quench. Eventually, the whole network of magnets can be damaged. This is avoided by quench detection and protection systems, e.g. Quench heaters \cite{Dahlerup-Petersen_1999aa} or CLIQ (Coupling-Loss Induced Quench), \cite{Ravaioli_2015aa}. Quench protection analysis can be further improved by coupling controller and circuit models \cite{Maciejewski_2017ab}.

		\section{Magnet Models}
		The electromagnetic field inside each magnet is described by Maxwell's equations \cite{Jackson_1998aa}, i.e., 
		\begin{align}
			\nabla\times\mathbf{E} &= -\frac{\partial \mathbf{B}}{\partial t}\;, &\quad
			\nabla\times\mathbf{H} &=  \frac{\partial \mathbf{D}}{\partial t}+ \mathbf{J}\;, &\quad
			\nabla\cdot \mathbf{D} &=  \rho \;, &\quad
			\nabla\cdot \mathbf{B} &= 0\quad
		\end{align}
		on a bounded domain $\Omega\subset\mathbb{R}^3$ and time interval $\mathcal{I}=(t_0,t_\text{end}]$ with specified boundary and initial 
		conditions on $\partial\Omega$ and $t_0$. They are interlinked by the material relations
		\begin{align}
			\mathbf{D} &= \boldsymbol{\varepsilon}\mathbf{E}\;,
			&\qquad
			\mathbf{H} &= \boldsymbol{\nu}\mathbf{B}\;,
			&\qquad
			\mathbf{J} &= \boldsymbol{\sigma}\mathbf{E} + \mathbf{J}_{\mathrm{s}}\;.
		\end{align}
		Here, $\mathbf{E}$ is the electric field, $\mathbf{D}$ the electric flux density, $\mathbf{H}$ the magnetic field, $\mathbf{B}$ the magnetic flux density, $\mathbf{J}$, $\mathbf{J}_\text{s}$ are the total, source current densities and $\boldsymbol{\rho}$ is the electric charge density. The positive (semi-)definite material tensors $\boldsymbol{\sigma}\succeq0$, $\boldsymbol{\varepsilon}\succ0$ and $\boldsymbol{\nu}\succ0$ define the electric conductivity, the electric permittivity and the magnetic permeability, respectively. For simplicity of notation, we consider only the case of linear material laws, see e.g. \cite{Cortes-Garcia_2017ab} for a generalization.
		
		Assuming a magnetoquasistatic approximation with $\frac{\partial}{\partial t}\mathbf{D}=0$ and
		introducing the magnetic vector potential $\mathbf{B}=\nabla\times\mathbf{a}$, the parabolic, semi-elliptic partial differential 
		equation
	\begin{align}
		\boldsymbol{\sigma}\frac{\partial\mathbf{a}}{\partial t}
		+
		\nabla\times(\nu\nabla\times\mathbf{a})
		= 
		\boldsymbol{\chi}{i}_\text{m}
		\label{eq:mqs}
	\end{align}
	is obtained, where the current density $\mathbf{J}_{\mathrm{s}}(\mathbf{r},t)=\boldsymbol{\chi}(\mathbf{r}){i}_\text{m}(t)$ is expressed in terms of scalar currents ${i}_\text{m}$. In practice, one often replaces the term $\boldsymbol{\sigma}\ \partial\mathbf{a}/\partial t$ by a homogenized model to avoid the resolution of small geometric details in the discretization, e.g. single strands within cables \cite{Bortot_2016aa}.
	Following \cite{Cortes-Garcia_2017ab}, equation \eqref{eq:mqs} is coupled via the voltage ${v}_\text{m}$ to a circuit using the flux linkage
	\begin{equation}
		{\Phi}(t)=\int_{\Omega}\boldsymbol{\chi}(\mathbf{r})\cdot \mathbf{a}(\mathbf{r},t) \ \mathrm{d} \Omega.\label{eq:mqs_coup}
	\end{equation}

	The temperature in the domain $\Omega_\text{s}$ can be obtained from the heat balance equation
	\begin{equation}
		\rho C_{\mathrm{p}}\frac{\partial T}{\partial t} - \nabla \cdot (k \nabla T) = P_{\mathrm{s}} + P_{\mathrm{Joule}},\label{eq:temp}
	\end{equation}
	with suitable boundary and initial conditions, where $\rho$ is the mass density, $C_{\mathrm{p}}$ the heat capacity, $k$ the thermal conductivity 
	and $T$ the temperature. The Joule losses $P_{\mathrm{Joule}}$ are defined as
	$P_{\mathrm{Joule}} = q_{\mathrm{flag}}\,\sigma^{-1} \|\mathbf{J}_{\mathrm{s}}\|^2$ 
	and are activated in case of a quench by the sigmoid-type activation function $q_{\mathrm{flag}}$ that depends on time, the 
	magnetic flux density $\mathbf{B}=\nabla\times\mathbf{a}$ and current density $\mathbf{J}_{\mathrm{s}}$.
	The power density $P_{\mathrm{s}}$ in $\Omega_\mathrm{s}$
	relates the heat equation with the field solution \cite{Cortes-Garcia_2017ab}.
In case of quenching, i.e., $q_{\mathrm{flag}}>0$, the superconducting coils have a resistance with the voltage drop
	\begin{equation}
	{v}_\text{m} = \frac{\mathrm{d}}{\mathrm{d} t}{\Phi}+\mathbf{R}_\text{t}{i}_\text{m}
	\text{ with }
	\mathbf{R}_\text{t}=q_{\mathrm{flag}}\int_{\Omega_\text{s}}\boldsymbol{\chi}^{\,\top}\sigma^{-1}\boldsymbol{\chi} \ \mathrm{d} \Omega. 
	\label{eq:temp_coup}
	\end{equation}
	The resistance $\mathbf{R}_\text{t}$ inherits the dependencies on $t$, $\mathbf{B}$ and 
	$\mathbf{J}_\text{s}$.
As temperature increases, thermal stress is caused which leads to displacement. The problem of linear elasticity reads
	\begin{align}
		\nabla\cdot(\mathbf{C}:\boldsymbol{\epsilon})=\mathbf{F}(T,\mathbf{J},\mathbf{B})
		\label{eq:elasticity}
	\end{align}
	where $\boldsymbol{\epsilon} = \frac{1}{2} \left(\nabla\mathbf{u}+(\nabla\mathbf{u})^{\!\top}\right)$ denotes the strain tensor and $\mathbf{u}$ the displacement vector. Boundary conditions still to be defined. The right hand side $\mathbf{F}$ depends on the thermal stress and Lorentz forces. In turn, the computed displacement deforms the computational domain $\Omega$.

	\section{Networks of Magnets}
	Let us assume that we have $N_\text{m}$ magnets connected within a network. Then each magnet $n=1,...,N_\text{m}$ is given by the same multiphysical model which is described by the coupled system of partial differential equations \eqref{eq:mqs}-\eqref{eq:elasticity} defined above. We address the solution of each model abstractly by the operator $\Psi_n$, such that
${v}_{\text{m},n} = \Psi_n({i}_{\text{m},n})$ ($n=1,...,N_\text{m}$), where the internal variables $\mathbf{a}_n$, $T_n$, $\mathbf{u}_n$ are suppressed as the electrical coupling to neighboring magnets is established only via currents ${i}_{\text{m},n}$ and voltages ${v}_{\text{m},n}$. Let us collect all solution operators as 
	\begin{align}
		\mathbf{v}_{\text{m}} = \boldsymbol{\Psi}(\mathbf{i}_{\text{m}})
		\label{eq:multiphysics}
	\end{align}
with currents $\mathbf{i}_\text{m}^{\!\top}=[{i}_{\text{m},1},{i}_{\text{m},2},...,{i}_{\text{m},N_\text{m}}]$ and voltages 
$\mathbf{v}_\text{m}^{\!\top}=[{v}_{\text{m},1},{v}_{\text{m},2},...,{v}_{\text{m},N_\text{m}}]$. Let an electrical network consist of simple 
elements and described by the modified nodal analysis \cite{Ho_1975aa}, i.e.,
	\begin{align}
		\label{eq:circuit_1}
		\mathbf{A}_\text{C}\mathbf{C}\mathbf{A}^{\!\top}_\text{C}\frac{\mathrm{d}}{\mathrm{d} t}\boldsymbol{\varphi}
		+\mathbf{A}_\text{R}\mathbf{G}\mathbf{A}^{\!\top}_\text{R}\boldsymbol{\varphi}
		+\mathbf{A}_\text{L}\mathbf{i}_\text{L}
		+\mathbf{A}_\text{V}\mathbf{i}_\text{V}
		\boxed{+\mathbf{A}_\text{m}\mathbf{i}_\text{m}}
		&= -\mathbf{A}_\text{I}\mathbf{i}_\text{s}(t)\;,\\
		\mathbf{L} \frac{\mathrm{d}}{\mathrm{d}t}\mathbf{i}_\text{L}-\mathbf{A}^{\!\top}_\text{L}\boldsymbol{\varphi}&=\mathbf{0}\;,\\
		\mathbf{A}^{\!\top}_\text{V}\boldsymbol{\varphi}&=\mathbf{v}_\text{s}(t)\;,\\
		\Aboxed{\mathbf{A}^{\!\top}_{\text{m}}\boldsymbol{\varphi}&=\mathbf{v}_{\text{m}}(t)}.
		\label{eq:circuit_end}
	\end{align}
	The position of each device in the network is encoded by a sparse incidence matrix $\mathbf{A_\star}$ which consists only of entries from the set 
	$\{-1,0,1\}$. The corresponding branch voltage is denoted by $v_\star= A^{\!\top}_\star \boldsymbol{\varphi}$ and the current through the device 
	is $i_\star$. The simplest network elements are inductors, capacitors and resistors; they are determined by lumped constitutive parameters, 
	collected in the matrices of capacitances $\mathbf{C}$, inductances $\mathbf{L}$ and conductances $\mathbf{G}$. Time dependent voltage and 
	current sources are given as vector-valued functions $\mathbf{v}_\text{s}(t)$ and $\mathbf{i}_\text{s}(t)$. The unknowns of the 
	system are the scalar potentials $\boldsymbol{\varphi}$ and a few currents, i.e. $\mathbf{i}_\text{L}$ and $\mathbf{i}_\text{v}$. More 
	complicated devices can be modeled by controlled sources, see e.g. \cite{Estevez-Schwarz_2000aa}. Finally, the terms enclosed in the two boxes 
	indicate the additional (non-standard) contributions due to the field/circuit coupling via equation \eqref{eq:multiphysics}.

\section{Iterative Coupling Accelerated by Reduced Models}
Instead of solving the large coupled system \eqref{eq:multiphysics}-\eqref{eq:circuit_end} monolithically, we propose an iterative scheme that fits 
into the framework of waveform relaxation or dynamic iteration, \cite{White_1985aa, Gander_2012aa}. This allows us to 
use a different discretization in time for each system and more importantly, even different software packages can be used.

Let us assume that we have an initial guess of the time-transient behavior of the voltages $\mathbf{v}_\text{m}^{(k)}(t)$ for $t\in\mathcal{I}$ and $k=0$. Then, in the simplest case, the following scheme can be performed
\begin{enumerate}
	\item solve the electric network equations \eqref{eq:circuit_1}--\eqref{eq:circuit_end} by considering the voltages $\mathbf{v}_\text{m}^{(k)}$ 
	to be known and extract the currents $\mathbf{i}_\text{m}^{(k+1)}$
	\item solve $n=1,...,N_\text{m}$ electrothermal problems \eqref{eq:mqs}-\eqref{eq:temp_coup} in parallel, each on domain $\Omega_n^{(k)}$ excited by the current ${i}_{\text{m},n}^{(k+1)}$. Extract voltages ${v}_{\text{m},n}^{(k+1)}$ and temperatures ${T}_{n}^{(k+1)}$.
	\item solve $n=1,...,N_\text{m}$ elasticity problems \eqref{eq:elasticity} in parallel, consider the temperature ${T}_{n}^{(k)}$ to be known, compute displacements $\mathbf{u}_{\text{m},n}^{(k+1)}$ and deform the computational domains $\Omega_n^{(k+1)}$.
	\item if not converged: increase counter $k=k+1$ and return to step 1.
\end{enumerate}
Due to the smallness of the displacements it is common practice to exclude the computation of the deformations from the iteration and solve the elasticity problem \eqref{eq:elasticity} only once in a post-processing step, i.e., after the iteration of steps 1-2 and 4 has converged. Convergence for the remaining electrothermal scheme is well understood and has been discussed based on Banach's fixed point theorem and an analysis of the contraction factor in \cite{Schops_2010aa},\cite{Cortes-Garcia_2017ab}.

The iterative method simulates each magnet in parallel but the computational burden might still be heavy. This can be mitigated by using reduced order models to decrease the number of iterations and eventually to reduce simulation costs, \cite{Rathinam_2002aa}. Let us assume that there is a reduced order model available that approximates the multiphysical problem \eqref{eq:mqs}-\eqref{eq:elasticity}. Then, the solution of the $n$-th magnet can be rewritten as 
	\begin{align}
		{v}_{\text{m},n}=\overline{\Psi}_n({i}_{\text{m},n})+\Delta{v}_{\text{m},n}
	\end{align}
where $\Delta{v}_{\text{m},n}$ is the deviation between the reduced model and the full one. This additional reduced model allows us to define a waveform relaxation in terms of voltage differences, i.e., 
	\begin{align}
		\mathbf{A}_\text{C}\mathbf{C}\mathbf{A}^{\!\top}_\text{C}\frac{\mathrm{d}}{\mathrm{d} t}\boldsymbol{\varphi}^{(k+1)}
		+\mathbf{A}_\text{R}\mathbf{G}\mathbf{A}^{\!\top}_\text{R}\boldsymbol{\varphi}^{(k+1)}
		+\mathbf{A}_\text{L}\mathbf{i}_\text{L}^{(k+1)}
		+\mathbf{A}_\text{V}\mathbf{i}_\text{V}^{(k+1)}
		\boxed{+\mathbf{A}_\text{m}\mathbf{i}_\text{m}^{(k+1)}}
		&= -\mathbf{A}_\text{I}\mathbf{i}_\text{s}(t)\;,\\
		\mathbf{L} \frac{\mathrm{d}}{\mathrm{d}t}\mathbf{i}_\text{L}^{(k+1)}-\mathbf{A}^{\!\top}_\text{L}\boldsymbol{\varphi}^{(k+1)}&=\mathbf{0}\;,\\
		\mathbf{A}^{\!\top}_\text{V}\boldsymbol{\varphi}^{(k+1)}&=\mathbf{v}_\text{s}(t)\;,\\
		\Aboxed{\mathbf{A}^{\!\top}_{\text{m}}\boldsymbol{\varphi}-\overline{\boldsymbol\Psi}({i}_{\text{m}}^{(k+1)})&=\boldsymbol{\Delta}\mathbf{v}_{\text{m}}^{(k)}}.
	\end{align}
This scheme still requires the simulation of the full model $\boldsymbol\Psi$ but may significantly reduced the number of evaluations.
On the other hand, sophisticated reduced models $\overline{\boldsymbol{\Psi}}$ may also increase the computational costs (particularly in the offline stage). However, in the simplest case the reduced model is given by an improved transmission condition in terms of the optimized Schwarz waveform relaxation \cite{Gander_2004aa}\cite{Cortes-Garcia_2017ab}. 

\section{Outlook}
The full contribution will show simulations of the mechanical response of a magnet during quench protection and magnet discharge. For CLIQ-protected magnets, there are two phenomena in the coil: an imbalanced current profile leading to inhomogeneous Lorentz forces and an almost homogeneous temperature profile \cite{Ravaioli_2015aa}. We quantify the impact of Lorentz forces and temperature gradients during a discharge as large stresses may lead to cable degradation over time.
 
\section*{Acknowledgements}
This work has been supported by the Excellence Initiative of the German Federal and State Governments and the Graduate School of CE at TU Darmstadt.


\begin{thebibliography}{10}

\bibitem{Bortot_2016aa}
L.~Bortot, M.~Maciejewski, M.~Prioli, A.~Fernandez~Navarro, S.~Sch\"{o}ps,
  I.~Cortes~Garcia, B.~Auchmann, and A.~Verweij.
\newblock Simulation of electro-thermal transients in superconducting
  accelerator magnets with comsol multiphysics.
\newblock In {\em Proceedings of the European COMSOL Conference 2016}, Munich,
  2016.

\bibitem{Cortes-Garcia_2017ab}
I.~Cortes~Garcia, S.~Sch\"{o}ps, L.~Bortot, M.~Maciejewski, M.~Prioli,
  A.~Fernandez~Navarro, B.~Auchmann, and A.~Verweij.
\newblock Optimized field/circuit coupling for the simulation of quenches in
  superconducting magnets.
\newblock {\em JMMCT}, 2(1):97--104, 2017.

\bibitem{Dahlerup-Petersen_1999aa}
K.~Dahlerup-Petersen, R.~Denz, J.~L. Gomez-Costa, D.~Hagedorn, P.~Proudlock,
  F.~Rodrinuez-Mateos, R.~Schmidt, and F.~Sonnemann.
\newblock The protection system for the superconducting elements of the large
  hadron collider at {CERN}.
\newblock In {\em Proceedings of the 1999 Particle Accelerator Conference},
  volume~5, pages 3200--3202, 1999.

\bibitem{Estevez-Schwarz_2000aa}
D.~Estévez~Schwarz and C.~Tischendorf.
\newblock Structural analysis of electric circuits and consequences for {MNA}.
\newblock {\em Int. J. Circ. Theor. Appl.}, 28(2):131--162, 2000.

\bibitem{Gander_2012aa}
M.~J. Gander.
\newblock Waveform relaxation.
\newblock In B.~Engquist, editor, {\em Encyclopedia of Applied and
  Computational Mathematics}. Springer, 2012.

\bibitem{Gander_2004aa}
M.~J. Gander and A.~E. Ruehli.
\newblock Optimized waveform relaxation methods for {RC} type circuits.
\newblock {\em IEEE Trans. Circ. Syst.}, 51(4):755--768, 2004.

\bibitem{Ho_1975aa}
C.-W. Ho, A.~E. Ruehli, and P.~A. Brennan.
\newblock The modified nodal approach to network analysis.
\newblock {\em IEEE Trans. Circ. Syst.}, 22(6):504--509, 1975.

\bibitem{Jackson_1998aa}
J.~D. Jackson.
\newblock {\em Classical Electrodynamics}.
\newblock Wiley and Sons, New York, 3rd edition, 1998.

\bibitem{Maciejewski_2017ab}
M.~Maciejewski, I.~Cortes~Garcia, S.~Sch\"{o}ps, B.~Auchmann, L.~Bortot,
  M.~Prioli, and A.~P. Verweij.
\newblock Application of the waveform relaxation technique to the co-simulation
  of power converter controller and electrical circuit models.
\newblock In {\em MMAR 2017}. IEEE, 2017.

\bibitem{Rathinam_2002aa}
M.~Rathinam and L.~R. Petzold.
\newblock Dynamic iteration using reduced order models: A method for simulation
  of large scale modular systems.
\newblock {\em SIAM J. Numer. Anal.}, 40:1446--1474, 2002.

\bibitem{Ravaioli_2015aa}
E.~Ravaioli.
\newblock {\em CLIQ -- A new quench protection technology for superconducting
  magnets}.
\newblock PhD thesis, University of Twente, 2015.

\bibitem{Schops_2010aa}
S.~Sch\"{o}ps, H.~De~Gersem, and A.~Bartel.
\newblock A cosimulation framework for multirate time-integration of
  field/circuit coupled problems.
\newblock {\em IEEE Trans. Magn.}, 46(8):3233--3236, 2010.

\bibitem{White_1985aa}
J.~K. White, F.~Odeh, A.~L. Sangiovanni-Vincentelli, and A.~E. Ruehli.
\newblock Waveform relaxation: Theory and practice.
\newblock {\em Trans. Soc. Comput. Sim.}, 2(1):95--133, 1985.

\end{thebibliography}
\end{document}